\documentstyle[prd,aps]{revtex}
\def\gtwid{\mathrel{\raise.3ex\hbox{$>$\kern-.75em\lower1ex\hbox{$\sim$}}}}
\def\ltwid{\mathrel{\raise.3ex\hbox{$<$\kern-.75em\lower1ex\hbox{$\sim$}}}}
\input epsf
\begin{document}
\draft
\title{Gravitational Lensing by Cosmic String Loops}
\author{Andrew A de Laix and Tanmay Vachaspati}
\address{Case Western Reserve University \\
Department of physics \\
Cleveland, OH 44106-7079}
\date{\today}
\maketitle
\begin{abstract}
We calculate the deflection of a light ray caused by the
gravitational field of a cosmic string loop in the weak field limit and
reduce the problem to a single quadrature over a time slice of the
loop's world sheet. We then apply this formalism to the problem
of gravitational lensing by cosmic string loops.  In particular, we
find an analytic solution for the special case of a circular loop
perpendicular to the optical axis. As examples of more complicated
loops, we consider two loops with higher frequency Fourier modes.  The
numerical analysis illustrates the general features of loop lenses.
Our estimates, using typical parameters for GUT scale loops, show that
the stringy nature of loop lenses can be observed for lensing systems
involving high redshift galaxies ($z \sim 2$), and we suggest that
gravitational lensing can confirm the existence of GUT scale strings
if they are the seeds for large scale structure formation.
\end{abstract}
\section{Introduction}
\label{intro}

Current research focuses on two scenarios for the formation of
structure in the universe: the first where structure formation was
seeded by adiabatic perturbations produced during an inflationary
epoch and the second where structure accretes around isocurvature
perturbations produced by topological defects such as cosmic strings,
global monopoles or textures. In the latter scenario it should be
possible to directly detect the presence of topological defects in the
present universe leading to immediate confirmation of the scenario. On
the other hand, the lack of direct evidence for topological defects in
the present universe can lead to constraints on the defect scenario
for structure formation and perhaps be considered as circumstantial
evidence in favor of the inflationary alternative. Thus it is quite
important to consider specific distinctive signatures of the various
topological defects that can be used to directly observe them.

Let us specifically consider cosmic strings, the model which will be
relevant to the work in this paper (for a review of cosmic strings,
see Ref. \cite{vilenkin}).  A number of observable features
produced by cosmic strings of mass density suitable for structure
formation have been discussed in the literature. These include
discontinuous patterns in the microwave background 
radiation \cite{bennett92},
generation of a gravitational wave background that could be detected by
noise in the millisecond pulsar timing \cite{barc} and gravitational 
lensing \cite{vil1,chrn,gott,paczynski,hindmarsh}. The
ongoing observations of anisotropies in the microwave background
radiation are expected to yield stronger constraints or positive
results over the next decade or so. The millisecond pulsar
observations can only impose tighter constraints on the string
scenario since a positive detection of gravitational waves does not
specifically imply the existence of cosmic strings.  
There has also been sporadic effort over the last decade to work
out the gravitational lensing signature of cosmic strings but, perhaps
due to the difficulties encountered in understanding the evolution of
the string network, no distinctive result emerged from these analyses.
However, an analytical framework for describing the string network has
been constructed over the last few years and the time seems ripe to
reconsider gravitational lensing as a tool for searching for
strings. The timing is also right from the observational viewpoint
since several new initiatives are underway that promise to survey much
wider and deeper regions of the sky.

In this paper we investigate gravitational lensing by cosmic string
loops\footnote{Gravitational lensing by global monopoles and textures
is likely to be less interesting since only a few of these are
expected to occur within our horizon. Also, their spherical symmetry
will lead to lensing that is harder to differentiate from that due to
conventional sources.}. We begin in Sec. II by estimating the
probability of string lensing and in so doing we review some of the
relevant properties of cosmic strings.  Our estimates are based on
recent results for the string network evolution summarized in
Ref. \cite{vilenkin}. Next, in Sec. III, we consider photon propagation 
in the metric of a loop. The problem appears to be quite difficult at
first because the oscillating loop is a complicated time dependent
gravitational source. Yet we are able to show that the problem reduces
to one that is static where a specific time slice of the loop's world
sheet is sufficient to determine the gravitational lensing effects. In
other words, the bending of light by a string loop is equivalent to
the bending of light by a static curved rod with non-uniform energy
density, a result similar to that for the energy shift of a photon
propagating in a string loop background first derived by Stebbins
\cite{stebbins}. We also rederive the energy
shift of the photon in the Appendix and recover a logarithmic term
that appears to have been eliminated by the regularization procedure
used in Ref. \cite{stebbins}.

Once we have set up the formalism for an arbitrary loop and described
some rudiments of gravitational lensing theory (Sec. \ref{basic}), we
apply it to treat the lensing due to a circular loop that
is oriented in a plane normal to the optical axis
(Sec. \ref{circlesec}). The results for the circular loop are in
agreement with the assumption in Ref. \cite{chrn} that the photons passing
through the loop remain undeflected. The deflection of a photon
trajectory not threading the loop can also be described quite simply
and the whole problem can happily be solved by hand without resorting
to numerical evaluation.

The perpendicular circular loop, however, is a very
special case as even a change in the orientation of the  loop
yields qualitatively different results, and loops with less symmetry
have completely different lensing behavior. The assumption that
photons passing through the loop remain undeflected fails for general
loops. We study the lensing due to several generic loops numerically
and provide image maps that promise to distinguish cosmic string
lensed images from more conventional gravitational lensing events
(Sec. \ref{numerical}). Here we also show that the Einstein radius of
the string loop is comparable to the typical loop size for any value
of the string tension and so the stringy nature of the loop plays a
crucial role in determining the structure of the lensed
images. Effective techniques --- for example, techniques that replace
the string loop by a point mass plus perturbations --- are unlikely to
yield successful approximations leading us to conclude that string
loop lenses ought to be observationally distinct from garden variety
astrophysical lenses. 
 
	In Sec. \ref{discussion} we summarize and discuss our main
results. We also qualitatively discuss the effects of long strings and
describe further work to come.  Finally, Sec. \ref{conclusion}
contains some concluding remarks.

\section{Lensing Probabilities with String loops}
\label{probability}

Gravitational lensing by cosmic string loops is an interesting problem
only if there is a realistic chance of observing a loop lens.  In this
section, we will estimate the typical size and number density of loops
and use these values to determine the likelihood of observing a string
loop. Our arguments will be based on scaling solutions to the string
network evolution which are indicated by numerical simulations and
semi-analytic treatments (for a review of scaling solutions see Ref.
\cite{vilenkin}).

The string network consists of two components --- the long (or
infinite) strings and the closed string loops --- assumed to have
formed during a phase transition in the very early universe. If the
strings are to seed large--scale structure formation, they should have
a linear mass density $\mu \sim 10^{22}$ gms/cm ( conveniently
expressed in Planck units as $G\mu \sim 2\times 10^{-6}$ where $G$ is
Newton's gravitational constant). At any epoch in the history of the
universe, curved sections of strings will oscillate under their own
force of tension; colliding and intersecting strings will undergo
reconnections; the strings will stretch under the influence of the
Hubble expansion, and the oscillating strings will lose energy
primarily to gravitational radiation. The complicated evolution of the
string network has been studied in a number of works and a consensus
is emerging that the long strings obey a scaling solution
\cite{bennett90,allen}. In other words, the energy density in the long
strings scales with time as $\rho_L \sim 1/t^2$ and the typical
distance between strings also scales with time, $L \propto t$, where
there are on the order of 10 long strings per horizon at any
epoch. The reconnections of long strings and large loops are found to 
copiously produce small loops. But there is generally less agreement 
on the precise 
number density of loops present at any epoch.  If the typical loop
produced is assumed to have a length $\ell \sim \alpha t$, then
numerical results (that do not take the gravitational back reaction into
account) place an upper limit of $\alpha < 10^{-3}$
\cite{bennett90,allen}.  To get a lower limit on $\alpha$, one
supposes that highly curved sections of string straighten out on very
short time scales due to gravitational radiation which occurs at a
rate
\begin{equation}
{{dE}\over {dt}} = \Gamma G \mu^2
\label{lossrate}
\end{equation}
where, $\Gamma$ is a numerical factor depending on the shape of the
loop and has been evaluated to be $\sim 60$ by considering several
classes of loops. Since the size of the loops is related to the
curvature of strings, the smallest loops that can be produced are
those that have lifetimes longer than the Hubble time scale. This
leads to $\alpha \approx \Gamma G \mu \sim 10^{-4}$ which we will use
for the purpose of numerical estimates. Then, for a scaling solution,
the number density of loops with length between $\ell$ and $\ell +
d\ell$ at a given time $t$ is
\begin{equation}
\label{loopdensity}
dn(\ell,t) \approx \frac{\nu d \ell}{t^2(\ell+\Gamma G \mu t)^2},
\end{equation}
where $\nu$ is about 0.5 \cite{vilenkin}.  Now we can estimate the
solid angle of sky coverage necessary to observe a string loop lens.
Using eq.\ (\ref{loopdensity}), the number of loops of length between
$\alpha t$ and $\beta\alpha t$ --- supposing that loops larger that
$\beta \alpha t$ will not make interesting lenses --- per solid angle
at redshifts less than $z_0$ is
%
%
\begin{equation}
\label{loopnumber}
\frac {dN_l}{d \Omega} = \int_{0}^{z_0} dz \int_{\alpha t}^{\beta \alpha
t} d\ell\frac{4}{H_0^3} \frac{dn(\ell,t)}{d\ell} \frac{1}{(1+z)^6}
\frac{( 1+z-\sqrt{1+z})^2}{\sqrt{1+z}},
\end{equation}
where $t = (2/3) H_0^{-1} (1+z)^{-3/2}$.  The various explicit factors
of the redshift arise from the changing volume element in an expanding
universe.  One can evaluate this integral analytically and find the
result
\begin{equation}
\label{loopnumber1}
\frac {dN_l}{d \Omega} = \frac{27}{2} \frac{ \nu \alpha (\beta - 1)
}{(\beta  \alpha + \Gamma G
\mu) ( \alpha + \Gamma G \mu)} \left[
\ln(1 + z_0) +z_0 -4 \sqrt{1 + z_0}  + 4 \right].
\end{equation}
For $z_0 = 1$, $\beta \sim 2$  and $\alpha =
\Gamma G \mu $, there are about 400 $(\Gamma G \mu_4)^{-1}$
loops per steradian which corresponds to about 1 loop for every
$(3^{\circ})^2 \Gamma G \mu_4$ of sky, where we have defined
$\Gamma G\mu_4 = \Gamma G \mu \times 10^4$.  But, this does not
necessarily tell us how many loop lenses we can expect to observe
because we have not yet factored in the odds that a galaxy will be
near enough to a given loop to be significantly lensed.  To resolve
this problem we require the luminosity distribution of high redshift
galaxies.

For an order of magnitude estimate of the luminosity distribution
of high redshift galaxies, we can use information about nearby galaxies 
and extrapolate to higher redshifts. Such an extrapolation is justified 
only if the effects of evolution can be ignored and we shall assume
that this is the case. The local luminosity function, $\phi$, is 
reasonably approximated by the Schechter function \cite{peebles},
\begin{equation}
\label{schechter}
\phi(u) = \phi_* u^\alpha e^{-u},
\end{equation}
where $u$ is proportional to the luminosity,
$\alpha = 1.07 \pm .05$, $\phi_* = 0.010e^{\pm .4} h^3~{\rm
Mpc}^{-3}$ and $h$ is the Hubble parameter in units of $100 {\rm
km~s^{-1}~Mpc^{-1}}$.  The number density of galaxies, $N_g$, per unit 
solid angle per unit redshift per unit flux is given
by
\begin{equation}
\label{ngalflux}
\frac{d^3N_g}{d\Omega dz d\kappa} = 4 \pi H_0^{-5}
z^4 \phi(\kappa z^2),
\end{equation}
where $\kappa$ is proportional to the flux.  For a flux limited or
equivalently a magnitude limited survey, we integrate over $\kappa$
from a lower bound to infinity.  The lower bound can be expressed in
terms of the apparent magnitude $m$ using the relation
\begin{equation}
\label{kappa}
\kappa = 9.0 \times 10^{16}h^{-2} 10^{M_*-m},
\end{equation}
where $M_* = -19.53 \pm .25 + 5\log h$ is the absolute
magnitude of a characteristic galaxy and a reasonable value for the
magnitude limit is $m = 24$.  For a flat universe with no cosmological
constant, age constraints suggest that we should use a value of $h =
0.5$ which gives a lower bound of $\kappa = 0.057$. We can express the
integral over the flux in terms of an incomplete gamma function which
gives us
\begin{equation}
\label{ngal}
\frac {d^2 N_g}{d \Omega d z} = 2.7 \times 10^8 z^2 \Gamma(\alpha+1,
0.057 z^2).
\end{equation}
Let us suppose that we have a loop located at about a redshift of one
away from us and we are observing galaxies that are about a
redshift of two to three. Then integrating eq.\ (\ref{ngal})
tells us that there are about $1.4 \times 10^9$ galaxies per steradian
to work with.  The typical angular separation between galaxies is
about 5.6 arc seconds, which we shall see in Sec. \ref{numerical} is
small enough that a foreground string loop can always be expected
to have a galaxy in its background. Therefore every loop has a good 
chance of being seen as a gravitational lens.

\section{Geodesic deflection by a string loop}
\label{deflection}

In this section, we consider the deflection of null
geodesics in the presence of a cosmic string loop, which in the limit
of geometric optics, will correspond to the photon paths. (Readers
only interested in lensing applications might wish to proceed directly
to the final result given at the end of the section.) We shall
assume that the background space--time is flat Minkowski space, and
ignore the effects of curvature or universal expansion. Since we
are interested in loops with radii much smaller that the horizon or
curvature scale, this approximation will be valid for determining the
photon deflection.  We shall further assume that the string
contribution to the metric is weak and may be treated as a
perturbation on the flat space.  Thus we may write the complete metric
as
%
%
\begin{equation}
\label{metricgen}
g_{\mu \nu} = \eta_{\mu \nu} + h_{\mu \nu},
\end{equation}
where we use the convention $\eta = {\rm diag}(-1,1,1,1)$, and $
h_{\mu \nu}$ is a small perturbation, so all terms of $O(h^2)$ will be
ignored. To fix the gauge, we apply the harmonic condition, that
is $g^{\mu \nu}\Gamma^{\lambda}_{\mu \nu} = 0$, which leaves us with a
simple wave equation for the metric
%
%
\begin{equation}
\label{wave}
\Box^2 h_{\mu \nu} = -16 \pi G S_{\mu \nu},
\end{equation}
where $S_{\mu \nu}$ is related to the stress--energy tensor by $S_{\mu
\nu}= T_{\mu \nu} - 1/2 \eta_{\mu \nu}T^{\lambda}_{\lambda}$.  It
should be noted that for a loop of finite size, the metric far from
the loop is asymptotically flat.

In the geometric limit, we can treat the trajectories of photons as
the null geodesics of the metric. If $\lambda$ is an affine parameter,
then the momenta are given by
%
%
%
\begin{equation}
\label{momentum}
P^{\mu} = {dx^{\mu} \over d \lambda}.
\end{equation}
The evolution of the $P^{\alpha}$ are determined by the geodesic equation
%
%
%
\begin{equation}
\label{geodesic}
{dP^{\alpha} \over d \lambda} + \Gamma^{\alpha}_{\mu \nu} P^{\mu}P^{\nu} = 0,
\end{equation}
with the constraint $P^{\mu}P_{\mu} = 0$.  In the weak field limit,
the Christoffel symbol, $\Gamma$, is
%
%
%
\begin{equation}
\label{christ}
\Gamma^{\alpha}_{\mu \nu} = {1 \over 2}\eta^{\alpha \delta}(h_{\mu \delta,
\nu} + h_{\nu \delta, \mu} - h_{\mu \nu, \delta}).
\end{equation}
To zeroth order, {\it i.e.}~in the absence of a loop, the Christoffel
symbol vanishes, implying the momenta are constants and the
coordinates grow linearly with $\lambda$.  In particular, one can
choose the affine parameter $\lambda$ to be $t/P^0$. 
Also, since the non--zero Christoffel components are all
first order, we can contract $\Gamma$ with the zeroth order momenta,
and further consider the Christoffel components to be functions of
only the zeroth order coordinates.  Let us be explicit by considering
the dimensionless four velocity which is defined by $\gamma^{\mu} =
P^{\mu}/P^0_{(0)}$ where $P^0_{(0)}$ is the zeroth order energy.  It
has a zeroth and first order part which we may write as $\gamma^{\mu}
= \gamma^{\mu}_{(0)} + \gamma^{\mu}_{(1)}$.  Rewriting eq.\
(\ref{geodesic}) using the appropriate substitutions, to first order
we get
%
%
%
\begin{equation}
\label{gammageo}
{d\gamma_{\alpha (1)} \over d t} = -(h_{\mu \alpha, \nu}-\frac{1}{2}
h_{\mu \nu, \alpha})\gamma^{\mu}_{(0)}\gamma^{\nu}_{(0)}.
\end{equation}

Let us now consider the light ray emitted from a static source residing 
at a distance much greater than the size of the loop and traveling to a 
distant observer, that is, we will consider our source and observer to lie 
effectively at infinity where spacetime is flat.  The components of 
$\gamma^{\mu}$ which appear explicitly on the right hand side
of eq.\ (\ref{gammageo}) are zeroth order, and the perturbation in the
metric $h_{\mu \nu}$ is a function only of the zeroth order
coordinates.  Therefore, we can integrate eq.\ (\ref{gammageo})
explicitly giving
%
%
%
\begin{equation}
\label{deflectint1}
\gamma_{\alpha(1)} = -\int_{-\infty}^{\infty}dt (h_{\mu \alpha, \nu}
-\frac{1}{2}h_{\mu \nu, \alpha})\gamma^{\mu}_{(0)}\gamma^{\nu}_{(0)}.
\end{equation}
However, $h_{\mu \alpha, \nu} P^{\nu} = dh_{\mu \alpha}/dt$ so that
the first term in the integral is just a vanishing surface term.  Thus
only the second term survives and we are left with
%
%
%
\begin{equation}
\label{deflectint2}
\gamma_{\alpha(1)} = \frac{1}{2}\int_{-\infty}^{\infty}dt~
h_{\mu \nu, \alpha}\gamma^{\mu}_{(0)}\gamma^{\nu}_{(0)}.
\end{equation}

We now require the metric produced by a string loop if we are
to calculate the geodesic deflection; in the weak field limit, this
problem has been solved.  The configuration of a string loop is
described by the position of the string $f^{\mu}(\sigma, t)$, where
$t$ is a time variable and $\sigma$ is a parameter along the loop.  The
equations of motion for the string in Minkowski space are given by
%
%
\begin{equation}
\label{eqofmotion}
\ddot{f}^{\mu} - f''^{\mu} = 0,
\end{equation}
with the constraints
%
%
\begin{equation}
\label{constraint}
\dot{f}^{\mu}f'_{\mu} = 0, \ \ \ \dot{f}^2 + f'^2 = 0 \ .
\end{equation}
Dots here refer to derivatives with respect to $t$ while primes
refer to derivatives with respect to $\sigma$. It is convenient
to choose,
\begin{equation}
f^{0} = t
\label{fo}
\end{equation}
and to write the string solutions
as a superposition of traveling waves with
non--linear constraints \cite{kibble}, namely,
%
%
\begin{equation}
\label{loopsoln}
{\bf f}(\sigma,t) = \frac{{\bf a}(\sigma-t)+{\bf b}(\sigma+t)}
{2}  \ , 
\end{equation}
\begin{equation}
{\bf a}'^2 = {\bf b}'^2 = 1 \ .
\end{equation}
For any closed loop of length $L$, solutions for {\bf f}
must satisfy the periodic condition
${\bf f}(\sigma + L, t) = {\bf f}(\sigma, t)$.  In the center of mass
frame of the loop, the functions ${\bf a}$ and ${\bf b}$ are periodic
as well, but this is not true in general. The description of the loop
as given in eq. (\ref{loopsoln}) also applies to loops 
that have a net momentum if we use the boundary
condition ${\bf a}(\sigma-t)-{\bf a}(\sigma-t+L) = {\bf
b}(\sigma+t+L)- {\bf b}(\sigma+t) = \bbox{\Delta}$, where 
$\bbox{\Delta}$ is the loop's center of mass velocity.

The energy-momentum tensor of the string is given in terms of 
$f$ by
%
%
%
\begin{equation}
\label{stress}
T^{\mu \nu} = \mu \int d\sigma (\dot{f}^{\mu}\dot{f}^{\nu}
- f'^{\mu}f'^{\nu})\delta^{(3)}({\bf x} - {\bf f}(\sigma,t)).
\end{equation}
With $T^{\mu \nu} $one can write the solution to eq.\ (\ref{wave}) as
an integral over a Greens function in retarded time, 
explicitly giving us
%
%
%
\begin{equation}
\label{metricsoln}
h_{\mu \nu}({\bf x},t) = 4 G \int d^3x' \frac{S_{\mu \nu}({\bf x}',\tau)}
{|{\bf x}- {\bf x}'|},
\end{equation}
where $\tau = t - |{\bf x}- {\bf x}'|$ is the retarded time.  Using the
stress--energy defined in eq.\ (\ref{stress}), we can evaluate the spatial
integral in eq.\ (\ref{metricsoln}),
%
%
%
\begin{equation}
\label{metricsoln1}
h_{\mu \nu}({\bf x},t) = 4 G \mu \int d \sigma 
\frac{F_{\mu \nu}(\sigma,\tau)} {|{\bf x}- {\bf f}|-({\bf x}- 
{\bf f})\cdot \dot{f}},
\end{equation}
where
%
%
%
\begin{equation}
\label{S}
F_{\mu \nu}(\sigma, t) = \dot{f}_\mu \dot{f}_\nu - f'_\mu f'_\nu
- \eta_{\mu \nu} \dot{f}^2.
\end{equation}
Let us use eq.\ (\ref{metricsoln1}) to test the validity the weak
field approximation.  Derivatives of $f_\mu$ are of order unity, so it
follows we may treat $F_{\mu \nu}$ as order one. The integral
over $\sigma$ gives a term of order $L$, the loop length. Roughly
then, the perturbation is small, {\it i.e.}~$h_{\mu \nu} \ll 1$, when
$|{\bf x}- {\bf f}| \gg G \mu L$.  The string energy density is given
by $ G \mu \ltwid 10^{-6}$, so that the weak field approximation will
break down only for photons which come within a small fraction of the
loop length of striking the string itself, implying that, for most lensing
scenarios, the approximation is valid.  There is, however, an
exception to this estimate in the case of cusps.  Cusps are points on
the loop which momentarily achieve the speed of light as the string
oscillates, and, if the velocity of the cusp points in the direction of the
photon trajectory, the deflection can be singular \cite{vachaspati}.
However, cusps only occur instantaneously, and the probability of a
photon encountering one may be neglected.

Eq.\ (\ref{metricsoln1}) might seem to be the appropriate starting point
for calculating ray deflections, but we shall see that it
is more convenient to solve this problem in Fourier rather than
coordinate space. To avoid confusion, let us use the following conventions,
%
%
%
\begin{eqnarray*}
F( x^\mu) &=& \frac{1}{(2 \pi)^4} \int d^4k e^{-i k_\lambda x^\lambda}
\tilde{F}(k^\mu), \\
\tilde{F}( k^\mu) &=&  \int d^4x e^{i k_\lambda x^\lambda}
{F}(x^\mu),
\end{eqnarray*}
and consider eq.\ (\ref{wave}) again. Transformed, it becomes
%
%
%
\begin{equation}
\label{Fourmet}
-k^{\lambda}k_{\lambda}\tilde{h}_{\mu \nu} = -16 \pi G \tilde{S}_{\mu \nu}.
\end{equation}
Now we choose the unperturbed photon trajectory to be 
$$
{\bf x} = {\hat \gamma} t
$$ 
and use
$\tilde{h}_{\mu \nu, \alpha} = -i k_{\alpha} \tilde{h}_{\mu \nu}$, 
to write
%
%
%
\begin{equation}
\label{timeint}
I_\alpha \equiv \int dt~h_{\mu \nu, \alpha}(t,\hat{ \gamma}t) =
-\frac{G}{\pi^3} \int dt \int d^4k~e^{-i t k_{\lambda}\gamma^{\lambda}}
i k_{\alpha} {\tilde{S}_{\mu \nu} \over k^{\lambda}k_{\lambda}},
\end{equation}
where all integrals are implicitly evaluated over an infinite
range. The integral over time can be evaluated, and just transforms
the exponential into a delta function $2\pi \delta(\hat {\gamma}\cdot
{\bf k} - k_{0})$.  If we now decompose the wave vector ${\bf k}$ into
components parallel and perpendicular to $\hat{\gamma}$ so that ${\bf
k} = (k_{\|},{\bf{k}}_{\bot})$, where ${\bf k} \cdot \hat{\gamma}
=k_{\|}$ and ${\bf{k}}_{\bot}$ is a two dimensional vector
perpendicular to $\hat{\gamma}$ then the delta function becomes
$\delta( k_{\|}- k_{0})$, which allows us to evaluate the integral
over $k_0$.  What remains is an integral over the three dimensional
vector ${\bf k}$
%
%
%
\begin{equation}
\label{deflect3}
I_\alpha = \frac{2 G}{\pi^2}\int dk_{\|}d^2k_{\bot} \frac{i k_{\alpha}
\tilde{S}_{\mu \nu}}{k_{\bot}^2},
\end{equation}
where $k_{\bot}$ refers to the magnitude of ${\bf k}_{\bot}$.  To make
any more progress, we need to evaluate $\tilde{S}$ in terms of ${\bf
f}$.  From eq.\ (\ref{stress}) one may infer that
%
%
%
\begin{equation}
\label{stresstilde}
\tilde{S}_{\mu \nu} = \mu \int d \sigma \int dt~F_{\mu \nu}(\sigma, t)
e^{i k_0 t}e^{-i {\bf k}\cdot {\bf f}(\sigma,t)}.
\end{equation}
One should note that {\bf f} is measured from an origin chosen somewhere
along the
zeroth order photon trajectory and {\it not} from the loop center of
mass.  Later we shall find it more convenient to decompose {\bf f}
into a component { \bf r} measured form the center of mass and ${\bf x_0}$
measured from the origin on the photon path, but for compactness we
stay with {\bf f} for now.  Replacing $k_0$ with $k_{\|}$, we can
substitute eq.\ (\ref{stresstilde}) into eq.\ (\ref{deflect3}) to get
%
%
%
\begin{equation}
\label{deflect4}
I_\alpha = \frac{2 G \mu}{\pi^2}\int d \sigma \int dt \int
dk_{\|}d^2k_{\bot} \frac{i k_{\alpha}}
{k_{\bot}^2}F_{\mu \nu}
e^{i k_{\|} (t - f_{\|})}e^{-i {\bf k_{\bot}}\cdot {\bf f_{\bot}}},
\end{equation}
where we have decomposed ${\bf f}$ into its parallel
and perpendicular components.

At this point it is necessary to choose $\alpha$. If we want to know
the redshift induced by the presence of a string, we would solve eq.\
(\ref{deflect4}) for $k_{\alpha} \rightarrow k_{\|}$; we will do this
in the appendix to compare our results with previous work, but for
now, we are interested in the deviation of the photon path from its
zeroth order direction. Thus we consider the components $k_{\alpha}
\rightarrow k_{\bot i}$.  In this case, one may perform the integral
over $k_{\|}$ to produce a delta function $\delta(t-f_{\|}(\sigma,t))
$, and so integrate out the time leaving
%
%
%
\begin{equation}
\label{deflect5}
I_{\bot i} = \frac{4 G \mu}{\pi}\int d \sigma \left [ \frac{F_{\mu \nu}}
{1 - \dot{f}_{\|}}
\int d^2k_{\bot} e^{-i {\bf k_{\bot}}\cdot {\bf f_{\bot}}}\frac{i k_{\bot i}}
{k_{\bot}^2}\right ]_{t = t_0},
\end{equation}
where $t_0$ is the solution to
%
%
\begin{equation}
\label{t0}
f_{\|}(t_0, \sigma) = t_0.
\end{equation}
The integration over ${\bf k_{\bot}}$ can be evaluated by recognizing
that $e^{-i {\bf k_{\bot}}\cdot {\bf f_{\bot}}}/ {k_{\bot}^2}$ is
proportional to the Fourier transform of the Greens function for the
two dimensional Laplacian operator, $-\pi \log (f_{\bot}^2)$. The
final result is suprisingly simple,
%
%
%
\begin{equation}
\label{deflectfinal}
I_{\bot i} = 8 G \mu \int d \sigma \left [ \frac{F_{\mu \nu}}
{1 - \dot{f}_{\|}} \frac {f_{\bot i}} {f_{\bot}^2}
 \right ]_{t = t_0}.
\end{equation}
Finally, if we define the two dimensional vector $\bar{\bbox{\alpha}}=
\hat{\gamma} _{\bot}(t \rightarrow -\infty)- \hat{\gamma}_{\bot}(t
\rightarrow \infty)$ to be the deviation of the photon from its zeroth
order trajectory, then from eq.\ (\ref{deflectint2})and eq.\
(\ref{deflectfinal}) we will get
%
%
%
\begin{equation}
\label{deviation}
\bar{\bbox{\alpha}} = -4 G\mu \int d \sigma \left [
\frac{F_{\mu \nu}(\sigma, t)
\gamma^{\mu} \gamma^{\nu}}
{1 - \dot{f}_{\|}} \frac {{\bf f}_{\bot}} {f_{\bot}^2}
\right ]_{t = t_0},
\end{equation}
where $t_0$ is determined by solving eq.\ (\ref{t0}).

\section{Gravitational Lensing}
\subsection{Basic Lens Theory}
\label{basic}

In the previous section we derived an expression for the deflection of
the photon momentum that passes near a cosmic string loop. Now we
would like to use this expression to determine how an intervening
string loop affects the images of various sources.  Let us define an
origin which is located at the center of mass of the loop, so the
vector ${\bf r(\sigma,t)}$ will trace the loop measured from this
origin.  Further, we shall define a second vector ${\bf{x}}_0$ which
points from the loop center of mass to the location of the photon at
$t = 0$, so we may rewrite ${\bf f}$ as the difference $ {\bf r} -
{\bf x}_0$.  Now let us define the optical axis as the line connecting
the loop center of mass to the observation point, that is ${\bf
r}_{\|}$ and ${\bf x}_{\|}$ both point in the direction defined by the
ray connecting the loop center of mass with the observer. Since the
source rays which are likely to be affected by the loop come in at
very shallow angles, we can treat them as rays parallel to the optical
axis for the purposes of calculating $\bbox{\alpha}$. We shall also
assume that the deflection occurs instantaneously in the lens plane
defined as the plane normal to the optical axis located at the loop
center of mass.  In each case, these approximations are accurate to
terms of order the loop size over the distance of the source or
observer to the lens (whichever is smaller).  For the loops we shall
be considering, these approximations will be more that adequate, since
typical distances are on the order of the horizon while the loop size
is about a factor $\Gamma G \mu$ smaller than that.

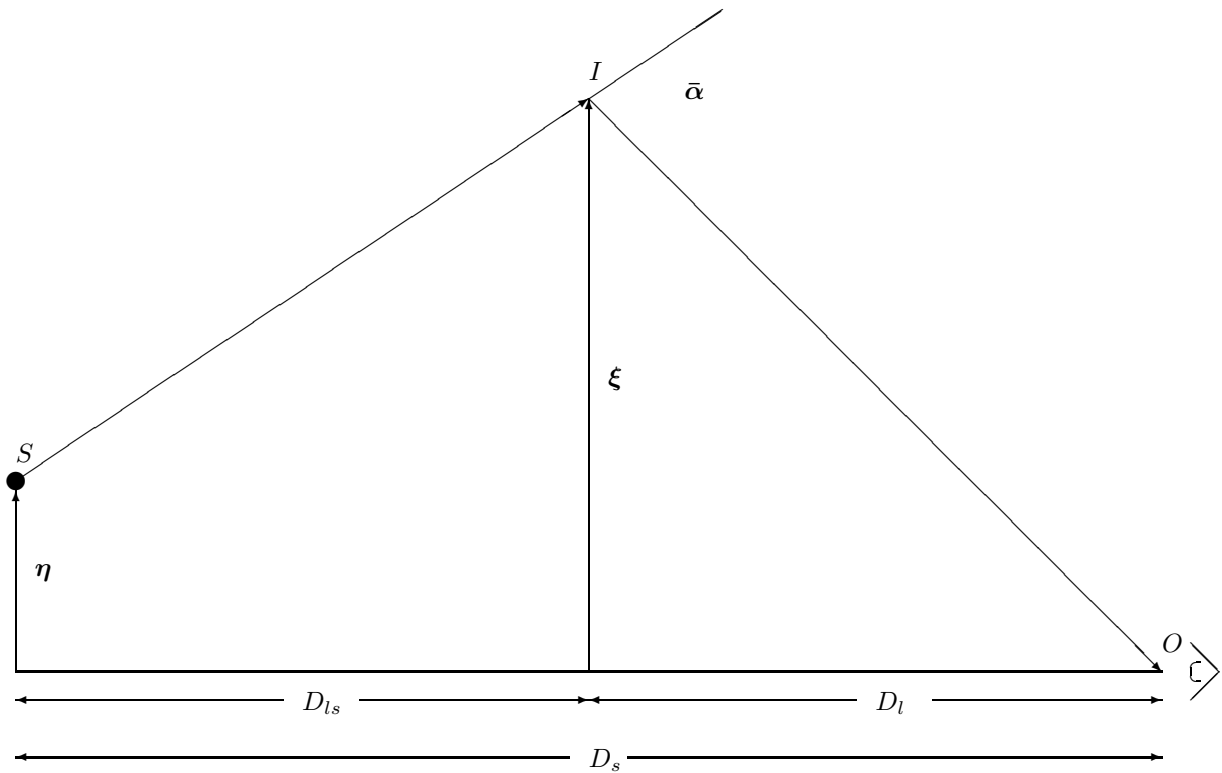
\begin{figure}[tbp]
\caption{\label{lensdiag}
A schematic representation of a gravitational lensing system.  The vectors
${\eta}$, ${\xi}$ need not be coplanar. The vector
$\bar{{ \alpha}}$ is defined as the difference of unit vectors
along $SI$ and $IO$.}
\setlength{\unitlength}{1in}
\vspace {1in}
\begin{picture}(6.5, 6)
\put(0, 3){\line(1,0){6}}
\put(0, 3){\vector(0,1){0.95}}
\put(3, 3){\vector(0,1){3}}
\put(0, 4){\vector(3,2){3}}
\put(3, 6){\vector(1,-1){3}}
\put(0,4){\circle*{.1}}
\put(.1,3.5){$\bbox{\eta}$}
\put(3.1,4.5){$\bbox{\xi}$}
\put(1.5,2.8){$D_{ls}$}
\put(1.4,2.85){\vector(-1,0){1.4}}
\put(1.8,2.85){\vector(1,0){1.2}}
\put(4.5,2.8){$D_{l}$}
\put(4.4,2.85){\vector(-1,0){1.4}}
\put(4.8,2.85){\vector(1,0){1.2}}
\put(3,2.5){$D_{s}$}
\put(2.9,2.55){\vector(-1,0){2.9}}
\put(3.2,2.55){\vector(1,0){2.8}}
\put(0,4.1){$S$}
\put(3,6.1){$I$}
\put(6,3.1){$O$}
\put(3,6){\line(3,2){.7}}
\put(3.5,6.){\bbox{\bar \alpha}}
\put(6.2,3){\oval(.1,.1)[tl]}
\put(6.2,3){\oval(.1,.1)[bl]}
\put(6.3, 3){\line(-1,1){0.15}}
\put(6.3, 3){\line(-1,-1){0.15}}
\end{picture}
\end{figure}

In Fig. \ref{lensdiag} we show a schematic representation of
the lensing system. A source lies in a plane perpendicular to the
optical axis a distance $D_{ls}$ from the lensing plane at the point
$S$.  The two dimensional vector in the source plane $\bbox{\eta}$
points from the optical axis to the source.  A ray emitted from the
source intersects the lensing plane at $I$, and the two dimensional
vector in the lensing plane $\bbox{\xi}$ points from the optical axis
to $I$.  At $I$, the ray is deflected by the vector
$\bar{\bbox{\alpha}}$. To be observed, the ray must intersect the
observer $O$ at a distance $ D_{l}$ from the lens plane on the optical
axis.  For small deflection angles, this condition is met if
%
%
\begin{equation}
\label{lenseeq}
\bbox{\eta} = \frac{D_s}{D_l} \bbox{\xi} - D_{ls} \bar{\bbox{\alpha}}
(\bbox{\xi}),
\end{equation}
where $D_s$ is the distance from the observer to the source plane.
This formula, referred to as the lens equation, gives us the location
of a source, $\bbox{\eta}$, given the location of its image,
$\bbox{\xi}$.  Usually we are interested in the inverse problem,
namely given a source location, where are its images located.  Note
that the inverse of eq.\ (\ref{lenseeq}) is not necessarily single
valued, so it is possible that a particular source has multiple
images.

We must digress for a moment to discuss exactly what is meant
by distance in the lens equation,  
for in an expanding universe, several definitions of distance are
possible. For example, if the universe were seeded with sources of a
known absolute luminosity, one could define the distance to any
of these sources to be proportional to the square root of the absolute
luminosity divided by the observed flux.  However, for the case of
gravitational lenses (Fig. 1), we are interested in the angular size of 
an image as it appears to a terrestrial observer, and we find that the
luminosity distance is not the best choice.  We should then like to
define the angular diameter distance, $d_A$, as the ratio of the 
object size and the angle that the object subtends at the location
of the observer. Then, using an FRW metric and assuming a flat, matter 
dominated universe with no cosmological constant, it is easy to show 
that the angular diameter distance is \cite{peebles}
%
%
\begin{equation}
\label{angdist}
d_A(z_1,z_2) = \frac{2}{H_0 (1+z_2)} 
\left[ {1 \over \sqrt{1+z_1}} - {1 \over \sqrt{1+z_2}} \right ].
\end{equation}
The lengths $D_s,~D_l$ and $D_{ls}$ are all angular diameter distances
given by this equation\footnote{A word of caution:
the real universe, especially at late times,
is clumped and only homogeneous on average.  A better distance choice
may come from models like the Dyer--Roeder equation \cite{dyer}, but
for the present we prefer not to consider these technical questions.}.

Given an extended source, not only can a lens change the
location and number of images observed, it can also affect the
magnification and shape as well.  To see this, let us, for the sake of
convenience, first redefine the lens equation in terms of
dimensionless variables. The loop length $L$ is defined in terms
of the energy of the loop as $E/\mu$. A convenient length scale for
us  is the loop radius which we define as $R \equiv L/2\pi$.  If we 
define
%
%
\begin{eqnarray}
\label{dimless}
{\bf x} &=& \frac{\bbox{\xi}}{R} \\
{\bf y} &=& \frac{\bbox{\eta}D_l}{D_s R} \\
\bbox{\alpha}({\bf x}) &=& \bar{\bbox{\alpha}}(\bbox{\xi})
\frac{D_{ls}D_l}{D_s R},
\end{eqnarray}
then the lens equation is reduced to
%
%
\begin{equation}
\label{lenseeqa}
{\bf y} =  {\bf x} - \bbox{\alpha}({\bf x}).
\end{equation}
One interesting question to ask, then, is given a narrow pencil beam
emitted by the source which subtends a solid angle $d \omega^* $, what is
the solid angle $d\omega$ subtended by its image?  This question is directly
related to the issue of magnification as the observed flux relative to the
emitted flux is just the ratio of the solid angles $ d \omega / d \omega^*$.
The answer may be found by looking at the Jacobian matrix
%
%
\begin{equation}
\label{jacobian}
 A_{ij} = \frac{\partial y_i}{\partial x_j},
\end{equation}
which gives us the magnification factor
%
%
\begin{equation}
\label{magnify}
\mu({\bf x}) =  \frac{1}{\det A({\bf x})},
\end{equation}
and the magnification is defined as the absolute value, $|\mu|$.  For many 
lensing systems, there will exist curves in ${\bf x}$ for which $\mu({\bf x})$
is infinite or equivalently the determinant of the Jacobian vanishes.
These curves are referred to as critical curves, and the
corresponding map of the critical curves into the source plane are
called caustics.  Caustics play an important role in determining the
number of possible images.  Consider a system for which the Jacobian
never vanishes.  The lens equation is then globally invertible and
therefore single valued.  Thus when there are no caustics, there can
be only one image of the source.  If there exists a caustic,
then the lens equation is only locally invertible and there may be
multiple images.  In fact there are many general theorems which can be
proved about caustics and images; we shall not discuss them here, but
the interested reader is directed to a detailed review of the subject
of gravitational lensing by Schneider, Ehlers and Falco
\cite{schneider}.

We have seen how a lens can affect the magnification of an
image, and we would now like to show how a lens can change the shape
of the observed image.  We shall restrict ourselves here to a
discussion of small sources so that we may treat the problem
differentially.  Consider two points on a source separated by a
distance ${\bf Y}$.  From the lens equation we can see, to first order
in the Taylor expansion, that the image displacement ${\bf X}$ will be
%
%
\begin{equation}
\label{distort}
{\bf X} =  A^{-1}{\bf Y}.
\end{equation}
Now consider a source that is a small circle.  From eq.\ (\ref{distort})
we can surmise that the image will be an ellipse with
major and minor axis pointing along the eigenvectors of $A^{-1}$
with lengths equal to their eigenvalues.  The most dramatic results
will appear near a critical curve.  Here, usually one of the
eigenvalues of $A^{-1}$ blows up, so that images are not only
magnified, but stretched as well.  What is observed is a stretched
image, but since the flux grows in proportion to the area, one observes
an image which appears to be a stretched version of the source but with 
the original brightness.

\subsection{An Analytic Example: The Perpendicular Circular Loop}
\label{circlesec}

In this subsection, we will consider a simple example of a
circular loop which lies in a plane perpendicular to the optical axis.  This
problem has the virtue of being analytically soluble, and it will also
illustrate some of the features that migh be expected for a more
general loop.  Most notably, there is a discontinuity in the
deflection $\bbox{\alpha}$ as light rays go from passing through
the loop to passing outside it which noticeably influences the
resulting images.

Let us begin by choosing the $z$ direction to point along the
optical axis.  For the perpendicular circular loop,
the configuration of the string is given by
%
%
\begin{equation}
\label{circleloop}
{\bf r} = R \cos({t \over R})[\cos({\sigma \over R}),\sin({\sigma \over R}),0],
\end{equation}
where $R$ is the maximal radius of the loop.  From eq.\ (\ref{S}), one
can see that for any planar loop perpendicular to the optical axis,
$\gamma^{\mu} \gamma^{\nu} F_{\mu \nu} = 1$.  Also, there is no parallel
component in ${\bf r}$, so $\dot{f_{\|}} = 0$.  Thus, using eq.\
(\ref{deviation}) and eq.\ (\ref{dimless}), we can write the
deflection as
%
%
\begin{equation}
\label{circledef}
\bbox{\alpha}({\bf x}) = -4 G \mu \frac{D_{ls}D_l}{ R D_s}
\int_{0}^{2 \pi} d \theta \frac{{\bf x}-
{\bf r}'(t_0)} {x^2 +r'^2(t_0) - 2 x r' \cos\theta},
\end{equation}
where ${\bf r}' = {\bf r(t_0)}/R$, $x\equiv |{\bf x}|$ and $t_0$,
defined by eq.\ (\ref{t0}), is a constant.  The behavior of this
integral is most easily seen by analytical continuation on to the
complex plane.  If we change variables to $z = \rho e^{i \theta}$,
then the integral in eq.\ (\ref{circledef}) is a contour integral
along the $\rho =1$ circle in the complex plane and hence reduces to a
sum of complex residues.  One finds that for image points such that $x
< r'$, the residues cancel and there is no deflection, but for image
points such that $x > r'$, the result is non--zero\footnote{This
discontinuous behavior will hold for all loops and is easily
understood in terms of the motion of poles in the complex plane.}. We
point this out because in principle the deflection for any loop can be
reduced to finding residues of a complex contour integral.  In
practice, however, the problem is complicated by the need to solve
eq.\ (\ref{t0}) to get $t_0$ as a function of $\sigma$, and there
appears to be no advantage in using the calculus of residues over
numerical integration to find $\bbox{\alpha}$, although there may be
other cases aside from the circular loop where this method could be
usefully applied.  Returning to the circular loop, one finds that
$\bbox{\alpha}({\bf x}) = 0$ if $|{\bf x}| < r'$ and
%
%
\[ \bbox{\alpha}({\bf x}) =  -8 \pi G \mu \frac{D_{ls}D_l}{R D_s}
  \frac{{\bf x}}{x^2} \ , \ \ \ 
{\rm if} \  |{\bf x}| > r' \ .
\]
For rays passing outside the loop, the deflection is exactly the same as
if the loop were replaced by a point mass.  Inside the loop, the ray is
undeflected, a feature which distinguishes a normal
circular loop from a point source.

Given the deflection, we will now consider the magnification,
specifically looking for the caustics.  The Jacobian matrix can
be written as $ A_{i j} = \delta_{ij}-\bbox{\alpha}_{i,j}$, so for the
circular loop we find that when $x > r'$,
%
%
\[ A = \left[
\begin{array}{cc}
1 - C({x_2^2 - x_1^2})/{x^4} & -2 C {x_1 x_2}/{x^4}\\
-2C {x_1 x_2}/{x^4} & 1 - C({x_1^2 - x_2^2})/{x^4}
\end{array} \right ] \]
where $C = 8 \pi G \mu {D_{ls}D_l}/{R D_s}$ and {\bf x} is given by
its components $(x_1,x_2)$. For $x < r'$,
$A$ is just the identity matrix and ${\rm det}(A) =1$. 
For $x> r'$, the determinant of $A$ is given by
%
%
\begin{equation}
\label{det}
\det(A) = 1 - \frac{C^2}{x^4} \ .
\end{equation}
One real root, namely $x = \sqrt{C}$, is proportional to the Einstein
radius $R_e = R \sqrt{C}$.  Note that this definition is consistent
with the familiar definition of the Einstein radius of a point mass,
namely $R_e =\sqrt{ 4 G M D_{ls}D_l/ D_s}$, where for a string the
mass is just $M = 2\pi R \mu$. So, if $\sqrt{C} > r'$, the loop lens
will have a circular critical curve with radius $\sqrt{C}$, but if
$\sqrt{C} < r'$ there will be no critical curve. In the latter case,
the loop would be difficult to detect because there will be only one
image of the source which will not be significantly distorted or
magnified.  In the former case ($\sqrt{C} > r'$) there are, strictly
speaking, two critical curves because the discontinuity at the
location of the string should be smoothed out on a scale given by the
thickness of the string.  Then the first critical curve is the usual
Einstein ring with radius $\sqrt{C}$ and the second occurs because the
determinant of $A$ must change from a positive value (namely 1) inside
the loop to a negative value just outside the loop (namely
$1-C^2/r'^4$) and hence ${\rm det}A = 0$ when the light ray passes
through the string\footnote{This second critical curve is of no
observable interest since the string thickness is a mere $10^{-30}$
cms or so.  It does have formal significance when considering lensing
theorems. In the following section, where we give image maps, we shall
ignore the effects of light rays that actually pass through the
string.}. For the case of the circular loop, we can analytically
invert the lens equation as well, and after a little algebra we find
%
%
\begin{eqnarray}
\label{image}
x_1 & = & \frac{y_1}{2} \left( 1 \pm \sqrt{1 + \frac {4C}{y^2}} \right) \\
x_2 & = & \frac{y_2}{2} \left( 1 \pm \sqrt{1 + \frac {4C}{y^2}} \right)
\end{eqnarray}
when $x > r'$, but ${\bf x} = {\bf y}$ when $x < r'$.  To help
visualize this rather complicated picture, we show in
Fig. \ref{circle} a series of images for a circular source as it is
displaced further away from the optical axis. We have chosen a loop
radius $r' = 1$ with a value of $C = 2.25$ corresponding to an
Einstein radius of 1.5.  It follows that there are two critical curves
with radii $r_c = 1$ and $1.5$

\subsection{Numerical Examples}
\label{numerical}

The perpendicular circular loop is one of the few cases which may be
treated analytically, but it is not a very generic loop.  In this
subsection, we would like to consider a few examples which represent
more general cases of cosmic string lenses.  We will consider loops
which are neither planar nor oriented in any special way with respect
to the optical axis and are characterized by reasonable astrophysical
parameters.

Suppose that we have a bright galaxy or cluster of galaxies at a
redshift of $z \sim 2$ which is lensed by a cosmic string loop.  Will
a ``typical'' loop produce any observationally distinctive images?
The answer to this question depends roughly on the value of the
Einstein radius of the system.  For a general loop, we define the
Einstein radius as
%
%
\begin{equation}
\label{einstein}
R_e \equiv \sqrt{8 \pi G \mu R \frac{D_{ls}D_l}{ D_s}},
\end{equation}
so if we were to replace the loop with a point source with mass equal
to the mass of the loop, the critical curve would be a circle with the
Einstein radius. If $R_e \gg R$ then the images which would be
significantly distorted by the string loop will appear to be similar
to those of an equivalent point source.  If $R_e \ll R$, then the loop
will not have a significant affect on any images of the source as most
of the mass of the loop will lie outside the Einstein radius. If
string loops are to produce distinctive images, we should have $R_e
\sim R$. So let us consider what may be reasonable parameters for a
loop lens system.

%
%
%
We can estimate the size of a ``typical'' string loop on the basis of
the string network evolution described in Sec. \ref{probability}.
Loops are expected to be produced from the network of long strings
with typical size $R \sim \Gamma G \mu t(z=1) /2 \pi $. These loops
gradually lose their energy to gravitational radiation
(eq. (\ref{lossrate})) but survive for about one Hubble period. (We
are assuming that the loops are not further fragmented significantly
by self-intersections. If that is the case, we would need to divide
the size estimate by the expected number of fragmentations and factor
in the survival period in the estimate of the lensing probability in
Sec. \ref{probability}.)  Then, using this value for the loop radius,
taking $\Gamma = 60$ and locating a source at $z = 2$ and the loop at
$z=1$, the Einstein radius is given by
\begin{equation}
R_e = R \sqrt{ {{16 \pi^2 } \over \Gamma}
{{D_{ls} D_l} \over {t D_s}}} \sim R \ 
\end{equation} 
--- which is in just the right range for distinctive string lenses.
Further note that the result $R_e \sim R$ is independent of the string
tension $\mu$ and so the stringy nature of loop lensing is important
for strings of any mass density.  We therefore expect that the images
produced by string loops will have characteristic features of stringy
lenses, thus offering the realistic hope that strings may be observed
definitively through gravitational lensing
\footnote{The wiggles on long strings can probably also be regarded 
as loops of size $R$ and hence the lensing by long strings would
mimick that of a linear array of loops. This would be another
characteristic feature of lensing by cosmic strings.}.

We shall consider two classes of loops to illustrate the kinds of
images that one may observe with cosmic string loops.  The first
is a mixture of the fundamental and the first excited mode 
%
%
\begin{equation}
\label{vloop}
{\bf r} = {R \over 2} \left\{ \sin \sigma_- {\bf i} + {1 \over 2}
\sin 2 \sigma_+ {\bf j} + [ \cos \sigma_- +{1 \over 2}
\cos 2 \sigma_+ ] {\bf k}\right \},
\end{equation}
where $\sigma_- = (\sigma - t)/R$ and $\sigma_+ = (\sigma + t)/R$.  We
will call this the ``two--loop'' for short since it is a superposition
of a fundamental mode with a frequency two mode.  The second loop
configuration is a class of loops found by Turok \cite{turok} which is
the general solution for a loop with frequency one and three Fourier
modes. This solution has the form
%
%
\begin{eqnarray}
\label{tloop}
{\bf r} & = &{R \over 2} \left \{
\left[ (1 - \alpha) \sin\sigma_- + {1 \over 3} \alpha
\sin3 \sigma_- + \sin\sigma_+  \right]{\bf i} \right . \\ \nonumber
& & -\left[ (1 - \alpha) \cos\sigma_- + {1 \over 3} \alpha
\cos 3 \sigma_- + \cos \phi \cos \sigma_+  \right]{\bf j} \\ \nonumber
& & \left . -\left[ 2 \sqrt{\alpha(1-\alpha)} \cos \sigma_- +
\sin \phi  \cos \sigma_+
\right]{\bf k}\right \}.
\end{eqnarray}
We shall call this the ``three--loop'' for short.  In general, the
coordinate system which defines $\{ {\bf i,~j,~k} \}$, {\it i.e.}~the
loop coordinates, can be rotated with respect to the optical axis,
which, for convenience, we shall define to be the $z$ axis.  The
relative orientation of the systems can be described by the three
Euler angles.  However, we are not particularly interested in the
orientation in the sky, so it is sufficient to consider only two of
the three. Thus, if we start with the systems $ \{ { \bf
x,~y,~z} \}$ and $\{ {\bf i,~j,~k} \}$ aligned, we shall first rotate
the loop in the $x-y$ plane from $x$ to $y$ through an angle
$\theta_1$.  Then we shall tilt the loop with respect to the observer
by rotating in the $y-z$ plane from $y$ to $z$ through an angle
$\theta_2$, giving us the final orientation of the loop coordinates
with respect to the optical coordinates.

To find the images for a given source is a numerically
non--trivial problem because one has to invert eq.\  (\ref{lenseeq})
to get ${\bbox \xi}$ as a function of ${\bbox \eta}$. This 
involves finding simultaneous roots of two equations in two
dimensions.
But, as there are no general methods to solve such a
problem, we are left with no choice but to map the entire image plane
(${\bbox \xi}$) into the source plane (${\bbox \eta}$) and then consider
the inverse mapping. (In practice, a lattice of points in the image
plane is mapped on to the source plane.)  To locate the images for a 
particular source point, we test
every triangle in the image plane that can be formed by three
neighboring points on the lattice.  If, when mapped to the source plane, the
image triangle encompasses a source point, then we know that the
source has an image located somewhere inside the image triangle.
Calculating $\bbox{\alpha}$ entails performing one numerical
integration, so the inversion process is reasonably cheap with regards
to cpu time and good resolution is possible.  The triangle search
method, however, has one failing when finding images with strings.
Since $\bbox{\alpha}$ will be discontinuous for light rays which pass
just to either side of the string, image triangles which are formed by
points on either side of the loop may encompass sources which they
would not if all the image points were on one side or the other of the
loop.  These are spurious images as they require the light rays to pass
through the string itself, so we have rejected these points when constructing
the images of specific sources (see the discussion in Sec. \ref{circlesec}). 

Let us now use this technology to generate some images for the loops
which we have discussed in this section\footnote{We will not
explicitly consider loops that have a net velocity though our
formalism applies directly to this case. We expect that the lensing by
moving loops will look very similar to the lensing by stationary loops
but that the loop will have a different effective shape due to the
condition in eq. (\ref{t0}).}.  In Fig. \ref{2loop2} we consider the
two--loop aligned with the optical coordinates so $\theta_1 = 0$ and $
\theta_2 = 90$.  The phase of the loop is given by $\psi$ which is
defined as the time at which the light ray intersects the lens plane
--- the plane normal to $\hat \gamma$ and containing the loop's center
of mass.  For this first example, $ \psi = 0$.  The source is located
at a redshift of two, the loop at a redshift of one and the loop
radius has been chosen to give an Einstein radius of $R_e/R = 1$ which
we have shown to be a typical value.  This choice corresponds to a
loop radius of $R = 2.3\times 10^{-5}~H_0^{-1}$.  The upper left
window shows the mapping onto the source plane of a series of vertical
lines spaced evenly at intervals of 0.02 $R$ in the image plane.  The
upper right panel shows the critical curves, the images of infinite
magnification, for this example.  The lower left panel shows the
unlensed images of a grid of source circles each defined by 15 points.
Finally, the lower right panel shows the images observed for the loop
lens along with the projection of the loop. (This field of sources is
primarily illustrative; realistic galaxies are larger compared to
realistic loops, and a subsequent set of figures will address this
issue.)  Images were found by the aforementioned method with a grid
resolution of $200\times 200$.  Figure \ref{2loop1} shows the same
two--loop now rotated through $\theta_2 = 0$, but now we only show the
lensed field and the critical curves.  Figure \ref{3loop1} shows the
same information as Fig. \ref{2loop1} but for a three--loop instead.
The loop parameters in this case were chosen to match those used by
Stebbins \cite{stebbins}: $\alpha = 0.5$, $\sin \phi = 0.5$, $\theta_1
= \theta_2 = 50$ and $\psi = 0$.  Finally, Fig. \ref{3loop2} shows the
same three--loop, but now at a phase $\psi = 120$.

In the next series of figures, we consider more realistic images that
might be observed with a loop lens and a distant galaxy.  Recall that
we expect the typical loop radius to be $\Gamma G \mu t/ 2 \pi$.
Taking $\Gamma = 60$ and locating the loop at a redshift of $z = 1$,
one finds that the loop subtends an angle of 3.2 $G\mu_6$ arc sec,
where $G \mu_6 \equiv 10^6 G \mu$. We have already stated that $G\mu_6
\sim 1$ if strings are responsible for seeding structure formation,
but we can more precisely constrain this value.  The best data comes
from the observation of the cosmic microwave background.  If strings
are responsible for the large scale temperature fluctuations observed
by COBE, then $G\mu_6 = 1.5 \pm .5$, as calculated by 
Bennett, Stebbins and Bouchet \cite {bennett92}. Below this range, 
strings cease to be interesting candidates for structure formation.  Upper
limits on the string tension have also been set by observing
millisecond pulsars.  The most recent data suggests that $G \mu_6 < 2$
\cite{barc}.  We consider $G \mu_6 = 1.25$ so that our loop diameter
subtends an angle of 4 arc seconds.  The angle subtended by the
visible portion of a distant galaxy at redshift $z \sim 2$, is roughly 1
arc second.  In Figs. \ref{b2loop1}-\ref{b3loop2}, we show several
examples of the images formed with a circular source with radius 1 arc
sec for each of our loop examples with $R = 2$ acr sec.  In each case
the Einstein radius was fixed to be one corresponding to a source
redshift of about two.  Source locations were chosen in these examples
primarily to show some of the more interesting features of the loop
lenses.

\section{Discussion}
\label{discussion}

The examples in the previous section may not represent an
exhaustive sample of possible loop configurations, but they do show
some of the generic features of loop lenses.  For both the two--loop
and the three--loop, we have considered configurations for which the
projected loop lies near the Einstein radius and more compact
configurations which reside well inside the Einstein radius.  For the
former case, Figs. \ref{2loop2}, \ref{3loop1}, \ref{b2loop2} and
\ref{b3loop1}, we notice that, like the perpendicular circular lens,
we can have images which pass through the center of the loop and lie
close to where the unlensed image would be.  This is perhaps the
distinguishing feature of loop lens images. In Figs.
\ref{b2loop2} and \ref{b3loop1} we see examples of relatively undistorted
images encircled by arcs and rings in a manner unlike what one would
expect for a more homogeneous lens mass distribution.  The more compact
loops produce, unsurprisingly, less spectacular images. In Fig.
\ref{b3loop2} there are some examples with three images which
may be distinguishable from ordinary lenses.  Still, for all loops, one
feature always differentiates them from ordinary lenses: namely that
the lens itself is dark. 

Would the observation of lensed images lacking an observed lens in
itself confirm the existence of cosmic string loops?  Unfortunately,
the answer is no; the existence of dark lenses in and of itself would
not be conclusive since there are other possibilities.  For example,
it has been suggested that dark matter could form halos without a
luminous component and these would result in dark lenses
\cite{hinshawkrauss,kandaswamy}.  Kandaswamy, Rees and Chitre
\cite{kandaswamy} found that a single dark matter halo is not
sufficiently dense to produce multiple imaging, but the alignment of
two such halos could.  Typical image separation was on the order of a
few arc seconds, similar to the string loop case. However, these halos
will have a distribution of matter consistent with collisionless
particles ({\it i.e.} an isothermal sphere), so their images should be
distinguishable from string loops.  A second possibility is that the
net effect of many distant lenses which individually would not produce
multiple images could add together to produce such an effect.  This,
however, has been shown to be statistically unlikely
\cite{schneider1}.  Perhaps the simplest explanation for a dark lens
is that it is not dark at all.  A cluster of galaxies could produce
two images which are separated on arc second scales, with a third
located typically an arc minute away.  The third image could easily go
unnoticed creating the false impression of a dark lens \cite{narayan}.
But again, the images should be distinctly different than those of a
string loop.  The good news then is that string loop lenses are
distinguishable from other dark lens possibilities, especially loops
which are not compact with respect to their Einstein radius.  Should a
dark lens be confirmed, detailed observations of the image structure
ought to determine the nature of the lens.

Another possibility that we have not examined in detail in this
paper but that is amenable to an identical analysis is the lensing
due to wiggly long strings. The presence of wiggles on long strings
means that the long string can probably be regarded as a sequence
of small loops. In this case, we should observe a linear sequence
of dark lenses with each lens having the characteristics of a string
loop lens. Such a lens is unlikely to occur in the context of
any other model.

\section{Conclusion}
\label{conclusion}

In this paper we have derived a method for calculating the deflection
of light rays due to the gravitational field of an oscillating string
loop.  We have shown that this problem can be reduced to an effective
static problem, greatly simplifying calculations.  The formalism was
then applied to the problem of gravitational lensing by cosmic
strings, and using typical loop parameters, we have shown that a loop
lens produces images on arc second scales, similar to galactic size
objects.  Specifically, we find that for $G \mu \sim (1-2)~10^{-6}$ ---
values consistent with structure formation, microwave background
anisotropies and millesecond pulsar timing limits --- strings can produce
images separated on arc second scales which would be observable by both
ground based telescopes and the Hubble space telescope and would 
have features that are distinctly different from other dark lens
candidates. This suggests that string loops can be definitively
observed as gravitational lenses. Furthermore, the lensing due to long
strings would appear like a linear sequence of lensings due to loops
and these would be the unmistakable fingerprints of cosmic strings.

\vfill
\eject

\appendix
\section{Effect of a string loop on the energy of a photon}
\label{cmbr}

The temperature fluctuations induced by cosmic string loops
has previously been investigated by Stebbins \cite{stebbins}, using
the harmonic gauge as we have done here.  Stebbins, however, used the
real space retarded Greens function to find the change in $\gamma_0$,
but, as we are solving the problem in Fourier space, it is useful to
compare our results.  Let us return then to eq.\ (\ref{deflect4}) and
replace $k_{\alpha}$ with $k_{\|}$ or equivalently $k_0$.  If we
rewrite the product $i k_{\|} e^{k_{\|}(t-f_{\|})}$ as $
1/(1-\dot{f}_{\|}) d/dt(e^{k_{\|}(t-f_{\|})})$, we can again evaluate
the $k_{\|}$ integral which now gives us
%
%
%
\begin{equation}
\label{temp1}
I_0 =
\frac{4 G \mu}{\pi} \int d\sigma \int dt \int d^2 k_{\bot} \frac{F{\mu \nu}}
{(1-\dot{f_{\|}})}\frac{e^{-i {\bf k_{\bot}}\cdot {\bf f_{\bot}}}}{k_{\bot}^2}
\frac{\partial}{\partial t}\delta(t-f_{\|}).
\end{equation}
The time integral can be evaluated by parts leaving
%
%
%
\begin{equation}
\label{temp2}
\frac{4 G \mu}{\pi} \int d\sigma \int d^2 k_{\bot} \left [
\frac{\partial}{\partial t}
\left (\frac{F{\mu \nu}}
{(1-\dot{f_{\|}})}\frac{e^{-i {\bf k_{\bot}}\cdot {\bf f_{\bot}}}}{k_{\bot}^2}
\right )\frac{1}{(1-\dot{f_{\|}})}\right ]_{t = t_0},
\end{equation}
and we can again evaluate the ${\bf k}_{\bot}$ integral as before.
Finally, the temperature change caused by the loop is given by
%
%
%
\begin{equation}
\label{temp3}
\frac{\Delta T}{T} = -4 G \mu \int d\sigma \gamma^\mu \gamma^\nu
\left [ \frac{\partial}{\partial t}\left (\frac{F{\mu \nu}}{(1-\dot{f_{\|}})}
\right )\frac{\log(f_{\bot}^2)}{(1-\dot{f_{\|}})}
+ \frac{2 F{\mu \nu}}{(1-\dot{f_{\|}})^2} \frac{\dot{ {\bf f}}_{\bot} \cdot
{\bf f}_{\bot}}{f_{\bot}^2}
\right ]_{t = t_0}.
\end{equation}
The appearance of the logarithm in the first term in the above expression may
be cause for concern as it appears to diverge for rays passing far from
the loop. However, Stebbins \cite{stebbins} has proven, using conservation of
energy that
%
%
%
\begin{equation}
\label{Tconstraint}
\int d\sigma \gamma^\mu \gamma^\nu
\left [ \frac{\partial}{\partial t}\left (\frac{F{\mu \nu}}{(1-\dot{f_{\|}})}
\right )\frac{1}{(1-\dot{f_{\|}})}
\right ]_{t = t_0} = 0,
\end{equation}
For rays far from the loop, {\it i.e.}~${\bf x}_0 \gg {\bf r }$, the
logrithm can be expanded to leading order, $\log(f_{\bot}^2) \approx
\log(x_0^2) -2 {\bf x}_0\cdot{\bf r }/x_0^2$, which falls like the
inverse of the distance and does not diverge.  

For completeness, we
sketch the proof of eq.\ (\ref{Tconstraint}).  Conservation of energy
requires that $\partial^{\nu}T_{\mu \nu} = 0$, so we can certainly
write
%
%
\begin{equation}
\label{bite}
\int d^3x \int dt~\delta({\bf x}\cdot \hat{\gamma}-t) \partial^{\nu}T_{\mu
\nu} ({\bf x},t) = 0.
\end{equation}
We are in essence performing the integral over the time slice as it
appears in eq.\ (\ref{temp1}).  Integrating the spatial components by
parts, one can easily verify that eq.\ (\ref{bite}) is equivalent to 
%
%
\begin{equation}
\label{me}
\int d^3x \int dt ~\delta({\bf x}\cdot \hat{\gamma}-t) \gamma^\nu
\frac{\partial}{\partial t} T_{\mu \nu} ({\bf x},t) = 0,
\end{equation}
given that $t = \hat{\gamma}\cdot {\bf x}$ is implied by the delta
function.  Using eq.\ (\ref{stress}) as a guide, we can write the stress
energy as 
%
%
\begin{equation}
\label{in}
T_{\mu \nu} = \mu \int d \sigma {\cal T}_{\mu \nu}(\sigma,t)\delta({\bf
x}-\bbox{f}(\sigma,t)),
\end{equation}
and substituting into eq.\ (\ref{me}) gives us
%
%
\begin{equation}
\label{my}
 \mu \int d \sigma \int d^3x \int dt \int dt' ~\delta({\bf x}\cdot
\hat{\gamma}-t)\gamma^\nu \frac {\partial}{\partial t}\left[{\cal
T}_{\mu \nu} (\sigma,t')\delta({\bf
x}-\bbox{f}(\sigma,t'))\delta(t-t')\right] = 0.
\end{equation}
We have added the extra delta function to enable us to see how to
evaluate this integral.  Let us replace it with a Fourier integral and
evaluate the derivative,
%
%
\begin{equation}
\label{big}
 \mu \int d \sigma \int d^3x \int dt \int dt' \int dk ~\delta({\bf x}\cdot
\hat{\gamma}-t)\gamma^\nu {\cal
T}_{\mu \nu} (\sigma,t')\delta({\bf
x}-\bbox{f}(\sigma,t'))\frac{i k}{2 \pi}e^{ik(t'-t)}  = 0.
\end{equation}
Now we can evaluate the integrals over $t$ and {\bf x},  leaving
us with 
%
%
\begin{equation}
\label{fat}
 \mu \int d \sigma \int dt' \int dk ~\gamma^\nu {\cal
T}_{\mu \nu} (\sigma,t')\frac{i k}{2 \pi}e^{ik(t'-f_{\|}(\sigma,t'))}  = 0,
\end{equation}
which is equivalent to (rewrite the product of $k$ and the exponential
as a time derivative)
%
%
\begin{equation}
\label{butt1}
 \mu \int d \sigma \int dt'\int dk ~\gamma^\nu \frac{ {\cal
T}_{\mu \nu}}{1-\dot{f_\|}(\sigma,t')}
\frac{\partial}{\partial t'}\delta(ik(t'-f_{\|}(\sigma,t')))  = 0.
\end{equation}
This can be integrated by parts, finally giving us
%
%
\begin{equation}
\label{butt}
 \mu \int d \sigma  ~\gamma^\nu \left[ \frac{\partial}{\partial t}
\left(\frac{ {\cal
T}_{\mu
\nu}}{1-\dot{f_\|}(\sigma,t)}\right)\frac{1}{1-\dot{f_\|}(\sigma,t)}
\right]_{t=t_0} = 0,
\end{equation}
where $t_0$ is again the solution to $f_\|(\sigma,t_0) = t_0$.  To
recover eq.\ (\ref{Tconstraint}),  one may contract the above with
$\gamma^\mu$ and recognize that $\gamma^\mu \gamma^\nu {\cal T}_{\mu \nu} = 
\gamma^\mu \gamma^\nu F_{\mu \nu}$.


We should point out that both terms in eq.\ (\ref{temp3}) 
are necessary for calculating the microwave background anisotropies
produced by string loops. In examples, however, we found that the contribution 
of the logarithmic term to the temperature fluctuation was only a few percent 
at most, and so we suspect that, in general, the logarithmic term 
that was overlooked in Ref. \cite{stebbins} will not make a significant
difference to the existing analyses \cite{bennett92}.

\newpage

%
%
%
\newpage
\begin{figure}[tbp]
\caption{\label{circle} The images resulting from a planar
circular loop of radius one (solid circle) for a series of circular
sources which spiral out from the origin.  Hatching indicates images
resulting from a particular source.  The dashed line shows the
Einstein radius selected for this example.} 
\end{figure}
\begin{figure}[tbp]
\caption{\label{2loop2} Lensing by a two--loop rotated through
$\theta_1 = 0$, $\theta_2 = 90$ and with phase $\psi = 0$.  Each panel
has dimensions $4R \times 4R$ where $R$ is the size of the loop.  The
upper left panel shows the mapping of a grid of lines in the image
plane onto the source plane.  The upper right panel shows the critical
curves in the image plane, {\it i.e.} the image points of infinite
magnification.  The lower left panel shows the unlensed images of a
grid of circles each defined by 15 points.  Finally, the lower right
panel shows the lensed images along with the projection of the loop at
time $t_0$.  }
\end{figure}
\begin{figure}[tbp]
\caption{ \label{2loop1}  The left
panel shows the projected loop and images of a field of sources like
those used in figure \protect\ref{2loop2}, but with a two--loop
rotated through $\theta_1 = \theta_2 = 0$ and phase $\psi =
0$.  The right panel shows the critical curve} 
\end{figure}
\begin{figure}[tbp]
\caption{\label{3loop1} Same as figure \protect\ref{2loop1}, but with
a three--loop rotated through $\theta_1 = \theta_2 = 50$ and phase
$\psi = 0$} 
\end{figure}
\begin{figure}[tbp]
\caption{\label{3loop2} Same as figure  \protect\ref{2loop1},
 but with a three--loop rotated through $\theta_1 = \theta_2 = 50$ and
phase $\psi = 120$} 
\end{figure}
\begin{figure}[tbp]
\caption{\label{b2loop2} The upper left panel shows the caustic of a
two--loop rotated through $\theta_1 = 0$, $\theta_2 = 90$ and phase $\psi
= 0$.  The other panels show the images as crossed points produced by a
circular source whose unlensed image is given by the filled points. 
Also shown is the caustic.}
\end{figure}
\begin{figure}[tbp]
\caption{\label{b2loop1} Same as figure \protect\ref{b2loop2}  but with
two--loop lens rotated through $\theta_1 = \theta_2 = 0$ and phase $\psi
= 0$.} 
\end{figure}
\begin{figure}[tbp]
\caption{\label{b3loop1} Same as figure \protect\ref{b2loop2} but with a
three--loop rotated through $\theta_1 = \theta_2 = 50$ and phase $\psi
= 0$. } 
\end{figure}
\begin{figure}[tbp]
\caption{\label{b3loop2} Same as figure \protect\ref{b2loop2} but with a
three--loop rotated through $\theta_1 = \theta_2 = 50$ and phase $\psi
= 120$.} 
\end{figure}

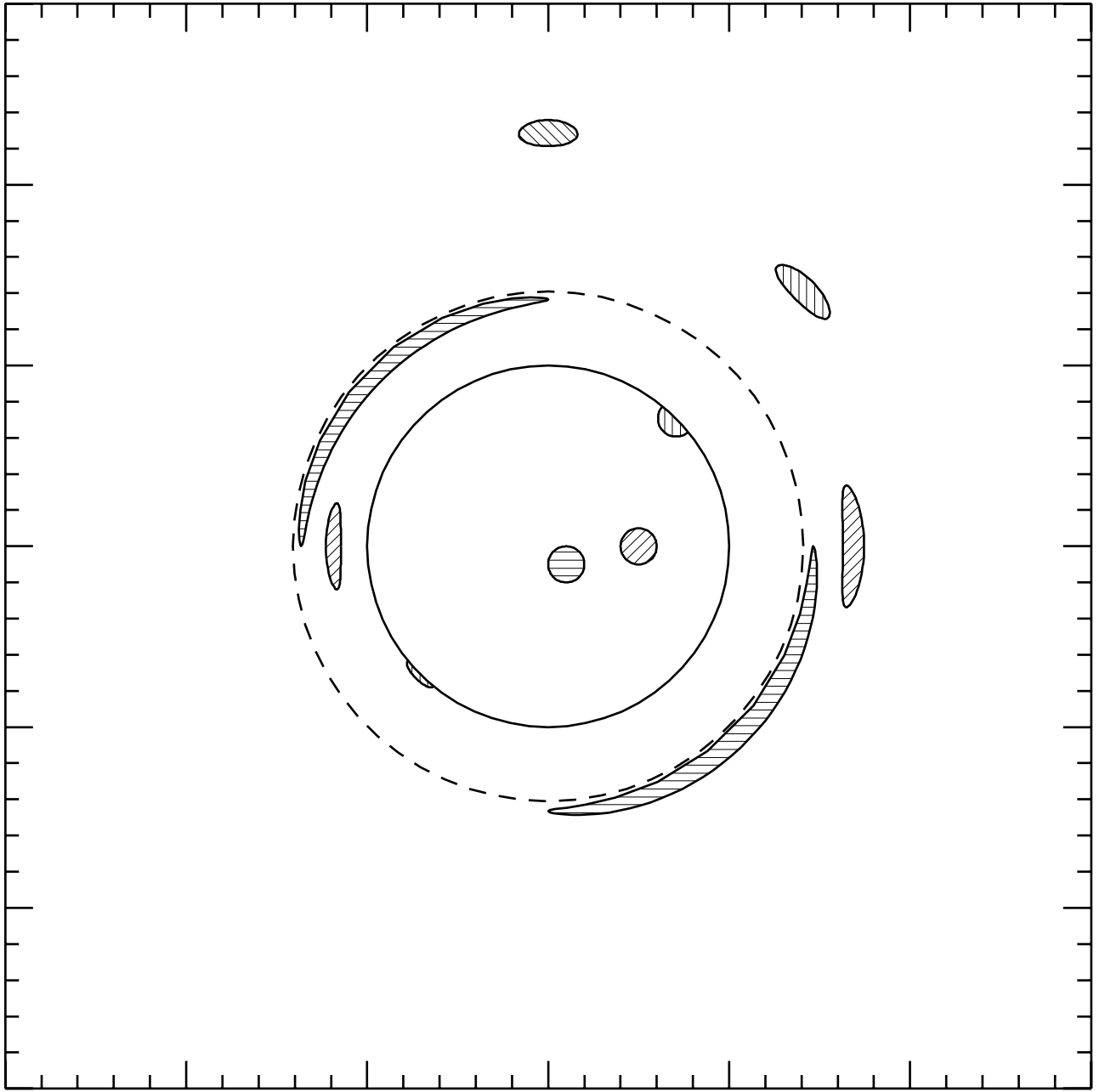
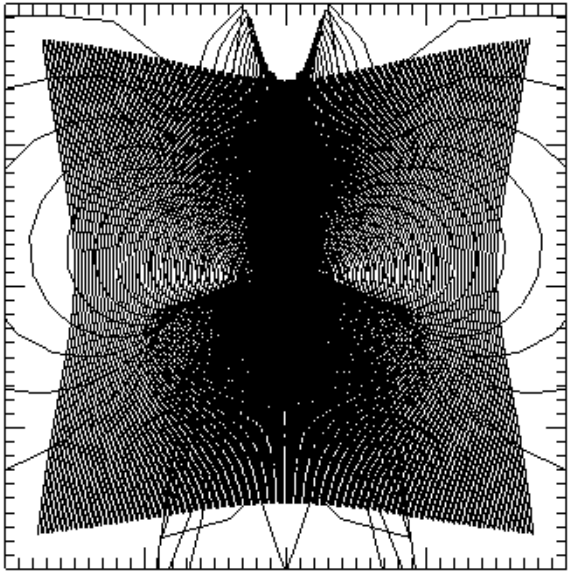
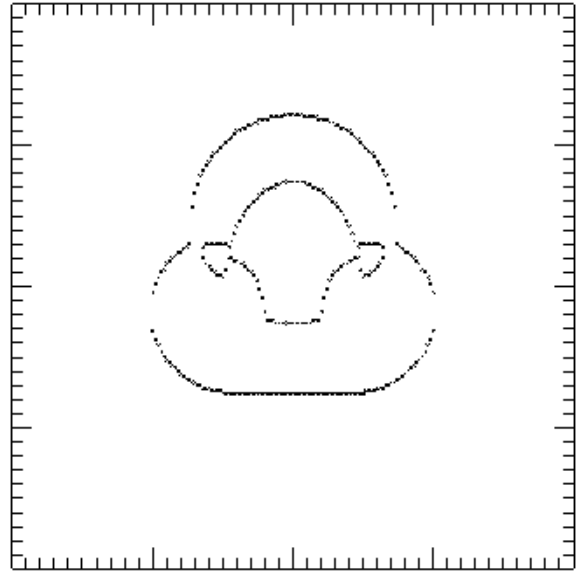
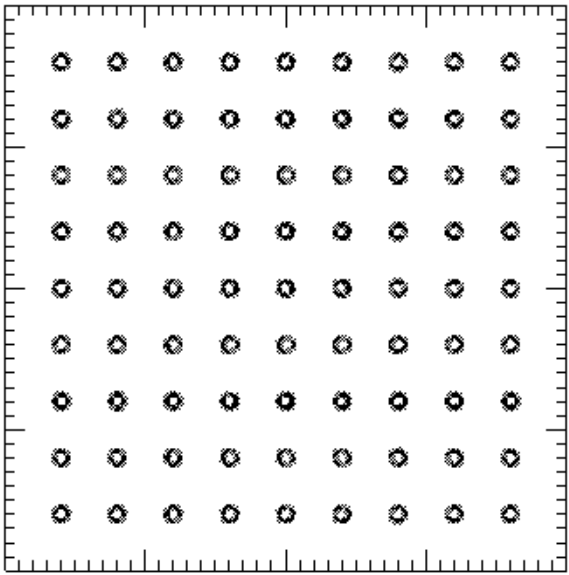
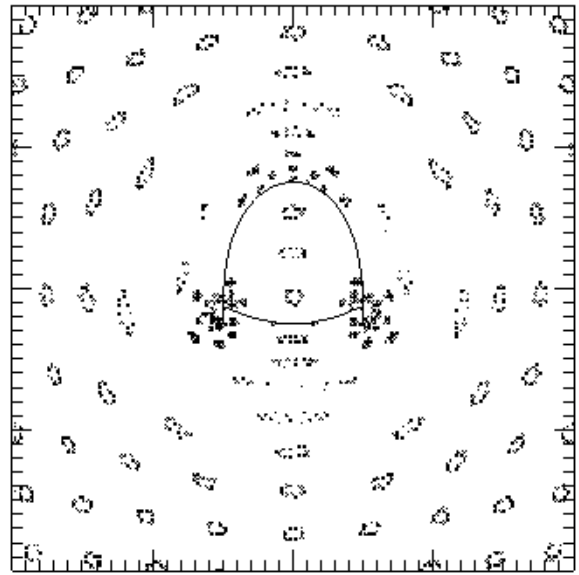
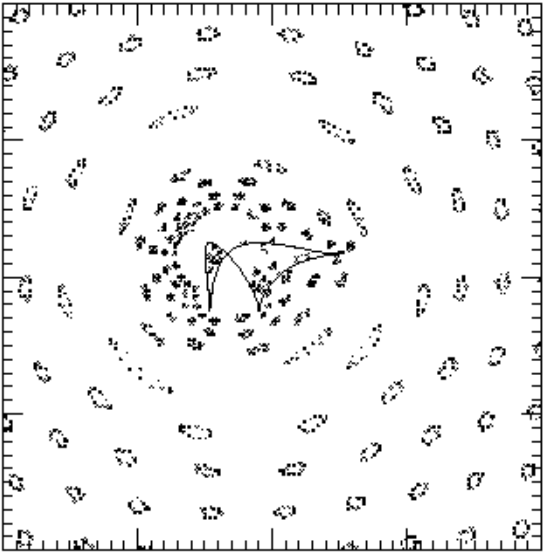
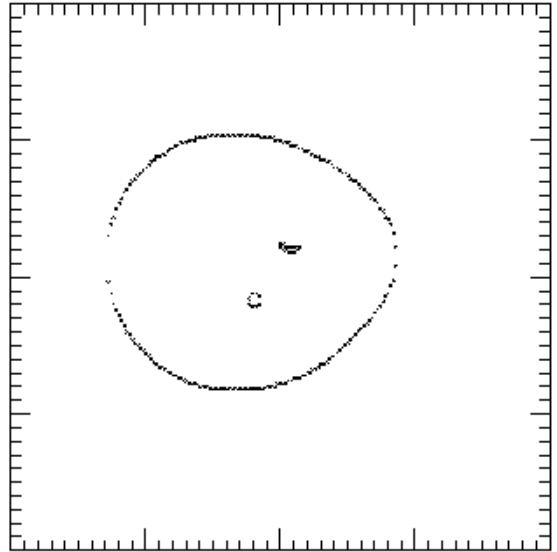
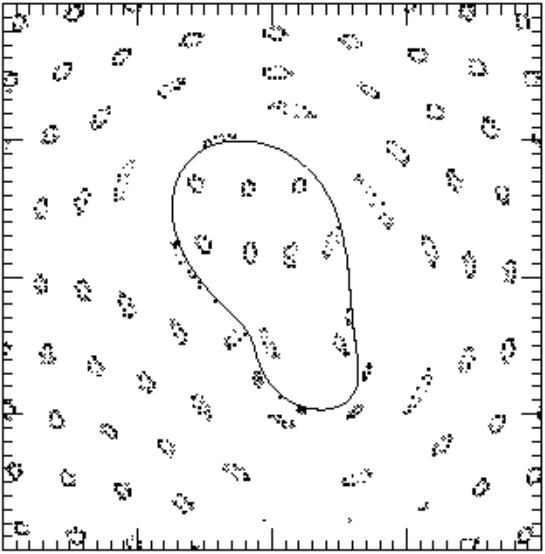
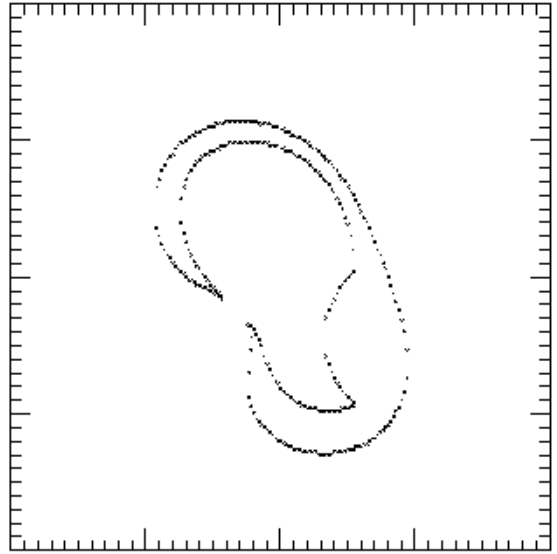
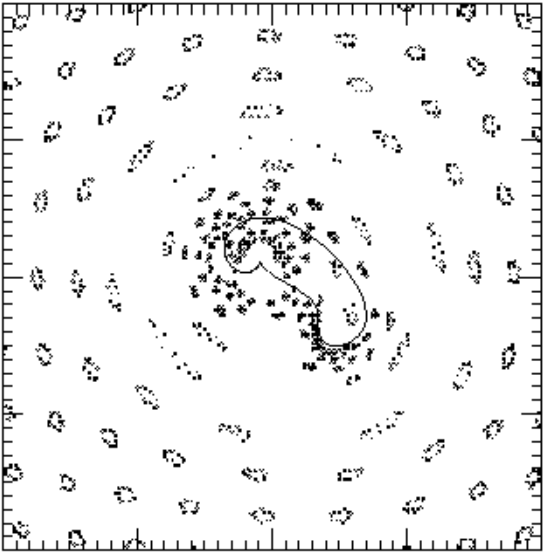
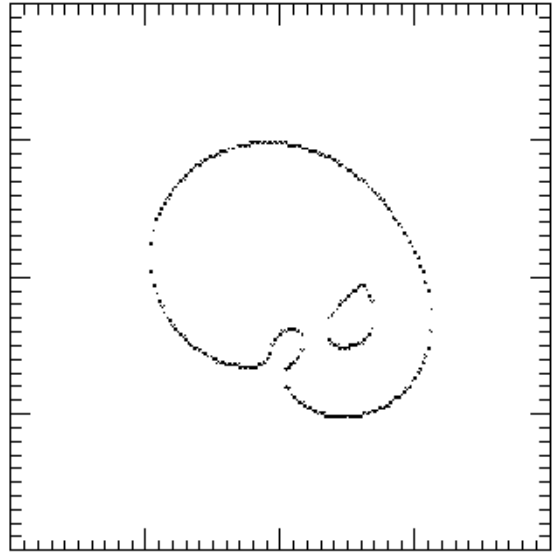
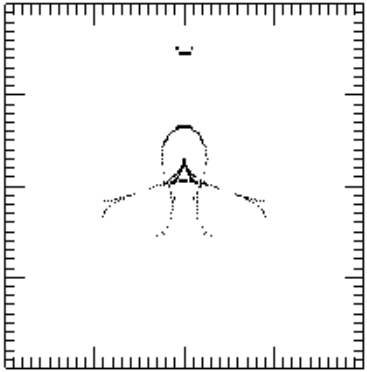
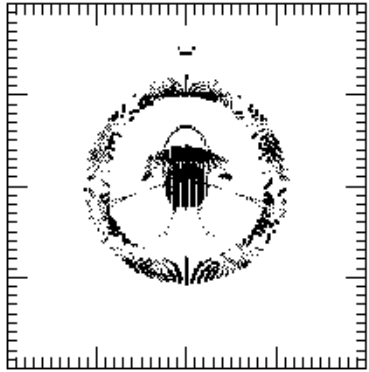
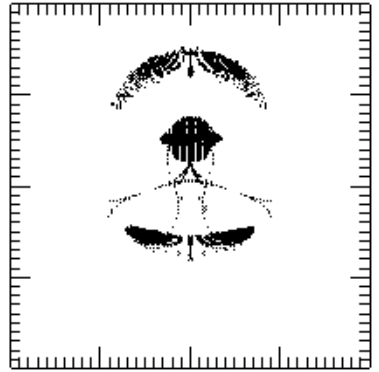
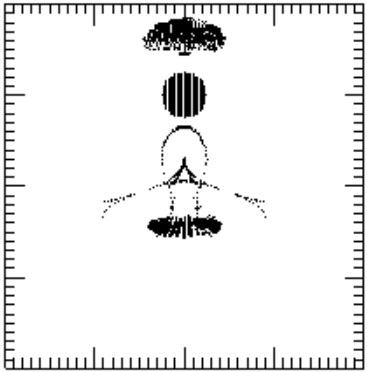
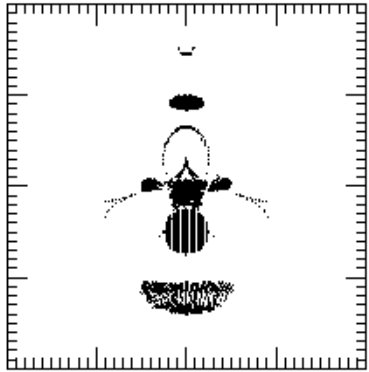
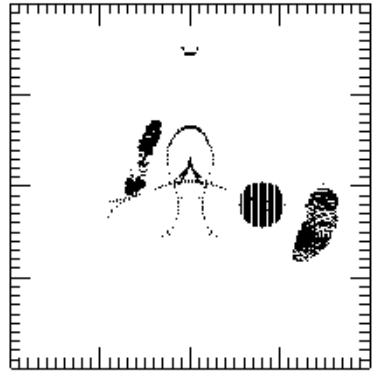
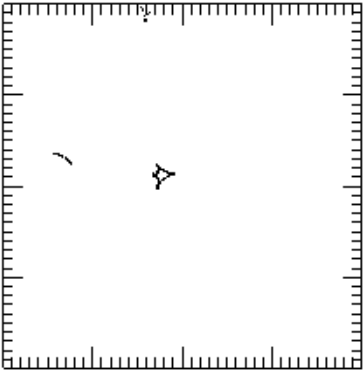
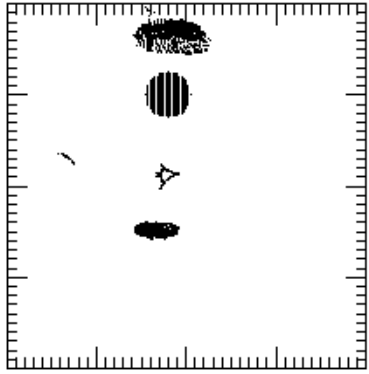
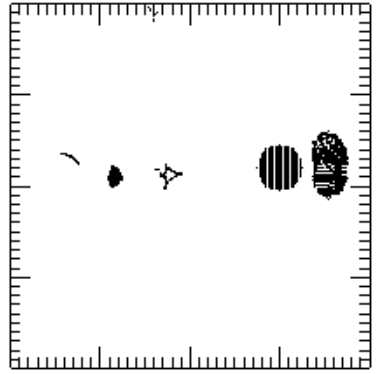
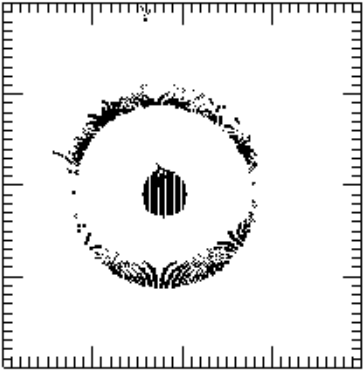
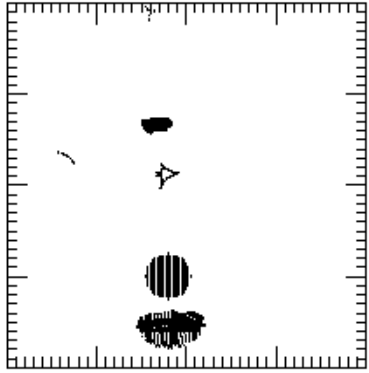
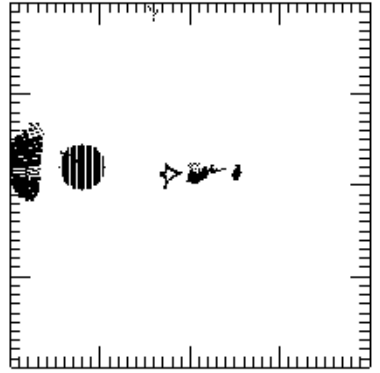
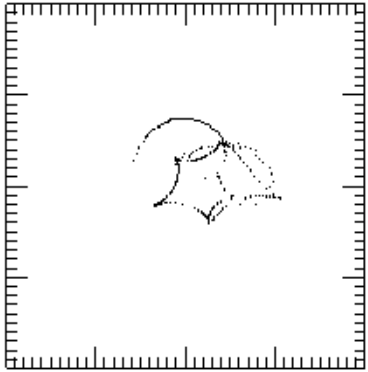
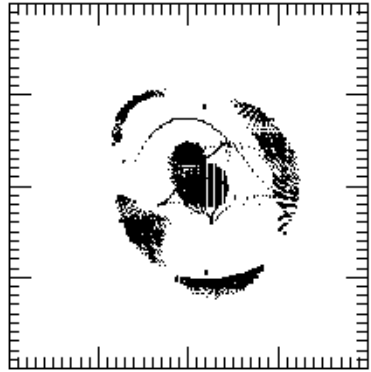
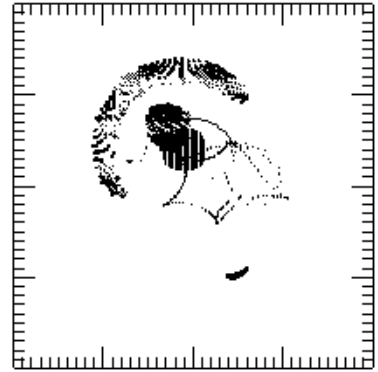
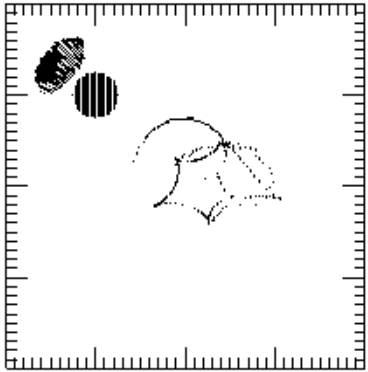
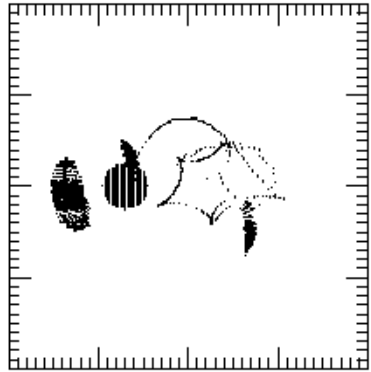
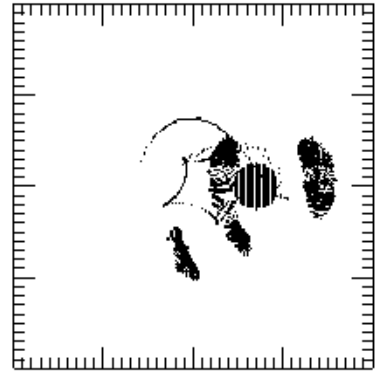
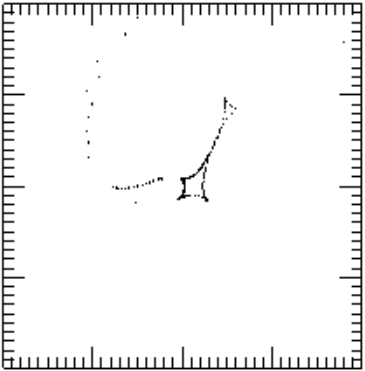
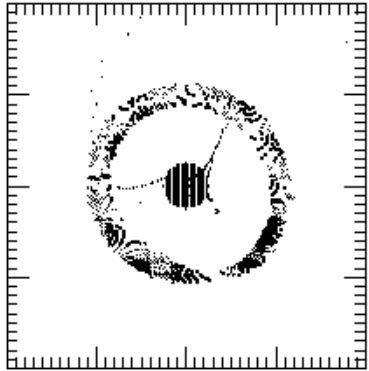
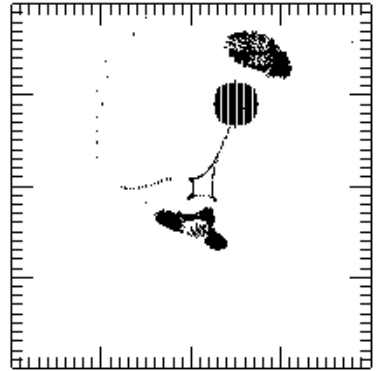
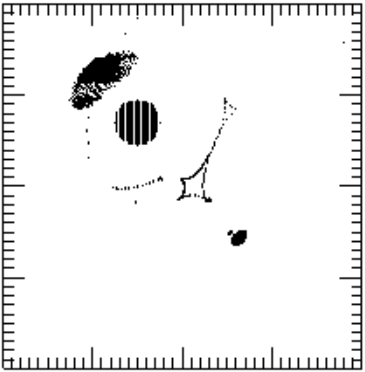
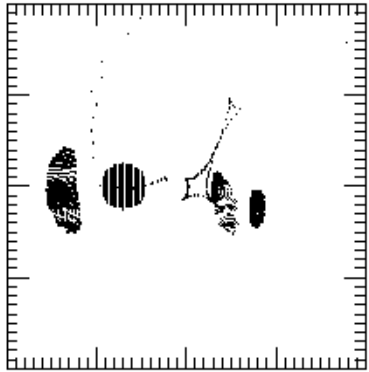
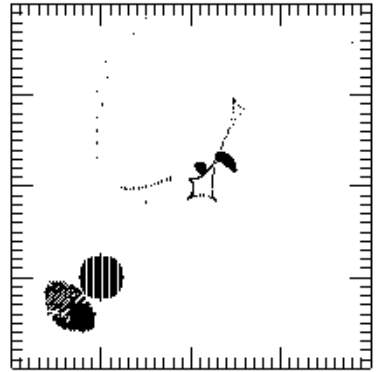
\end{document}